\documentclass{mn2e}
\usepackage{psfig}

\def\ltsima{$\; \buildrel < \over \sim \;$}
\def\lsim{\lower.5ex\hbox{\ltsima}}
\def\gtsima{$\; \buildrel > \over \sim \;$}
\def\gsim{\lower.5ex\hbox{\gtsima}}

\begin{document}

\title[Photoionization in dusty media III]{Time-Dependent 
Photoionization in a Dusty Medium. \\ III: The effect of dust on the
photoionization of metals}

\author[Lazzati \& Perna]
{Davide Lazzati$^1$ \& Rosalba Perna$^{2,3}$ \\ 
$^1$ Institute of Astronomy, University of Cambridge, Madingley Road,
Cambridge CB3 0HA, England \\
$^2$ Harvard Society of Fellows, 78 Mt. Auburn Street, Cambridge, MA
02138\\ 
$^3$ Harvard-Smithsonian Center for Astrophysics, 60 Garden Street,
Cambridge, MA 02138 \\
{\tt e-mail: lazzati@ast.cam.ac.uk; rperna@cfa.harvard.edu}
}

\maketitle

\begin{abstract}
We use the time-dependent photoionization and dust destruction code
developed by Perna \& Lazzati (2002) to study the time evolution
of the medium in a dusty gaseous cloud illuminated by a bright
central source that sets on at time zero. We study the case of a
bright source, which lasts for a time scale much smaller than the
recombination and dust creation time scales. For this reason an
equilibrium is never reached. We show that the presence of dust and
its properties, such as its composition, can have a big effect on
the time scale for the evaporation of the soft X-ray absorption, in
particular for ionizing sources with hard spectra. We discuss the
profile of evaporation of the soft X-ray absorbing column, as well as
how the apparent ionization state of the cloud evolves in time. We
finally consider the apparent metallicity of the cloud that is left
behind as a function of the cloud and ionizing source properties.
\end{abstract}

\section{Introduction}

The measurement of the amount of absorption in the soft X-ray band is
highly informative about the amount and physical state
(e.g. temperature, ionization parameter) of the material that lies
along the line of sight between a source and an observer. Observed
spectra are usually analyzed under the assumption that the absorber is
in equilibrium, either a thermal or photoionization equilibrium
(Morrison \& McCammon 1983; Done et al. 1992; Zdziarski et
al. 1995). This assumption is justified if the source that is studied
is constant or if its variability time scale is much longer than the
electron recombination time scale of the absorbing material. There are
however many variable sources in the universe that do not fulfill this
condition. Among them it is worth mentioning X-ray transients (see
e.g. Schwarz 1973), variable AGNs (see, e.g. Risaliti, Elvis \&
Nicastro 2002) and especially gamma-ray bursts (Perna \& Loeb 1998;
Lazzati \& Perna 2002, hereafter LP02). In all these cases, out to a
certain distance, photoionization acts as a destructive mechanism:
each time a photon is absorbed an electron is stripped, and therefore
the state of the absorber and its absorption properties are modified.

LP02 (see also Schwarz 1973) analyzed by means of numerical
simulations the absorption properties of a gaseous medium suddenly
illuminated by a strong photon source with a power-law spectrum. They
followed the ionization state of the 12 astrophysically relevant
elements and hydrogen producing time resolved transmittance spectra to
be compared with the data. In many cases, however, assuming that all
the interstellar medium (ISM) is in a gaseous phase is not
accurate. There can be molecules (that do not have sizable affect on
the X-ray opacity of the medium, see Perna, Lazzati \& Fiore 2002,
hereafter Paper II) and, most importantly, there can be dust
grains. These in particular are relevant to X-ray absorption since a
large fraction of heavy metals can be condensed in silicate dust
grains (Savage \& Mathis 1979; Laor \& Draine 1993). The presence of
dust grains in the ISM modifies the X-ray absorption properties in two
main ways. First, the presence of dust increases the opacity for UV
photons, that cannot therefore contribute to the ionization of metals
by stripping their external electrons. Most importantly, the metals
contained into a dust grain are shielded to photoionization, so that
they can survive in an almost neutral state for a much longer time
than the same atom in gaseous phase.

We (Perna \& Lazzati 2002, hereafter Paper I) have developed a code
that computes the time-dependent absorption properties of a dusty ISM
(possibly enriched with $H_2$ molecules) as a function of time under
the illumination of a power-law ionizing continuum that evaporates
dust grains, dissociates molecules and ionizes atoms. The code and the
physical processes considered are fully described in Paper I. The
properties of absorption in the optical range are analyzed in a second
paper (Paper II). Here, we focus on the properties of the absorber in
the X-ray range, and on how the presence of dust influences the
evolution of the opacity under different conditions.

In \S~2, we briefly introduce the code and describe the outputs that
will be used here, in \S~3 we discuss the main effects of the presence
of dust while in \S~4 we study how these effects depend on the
spectrum of the central ionizing source. In \S~5 we study how the
residual column density can be quantified with a single parameter and
in \S~6 we consider the shape of the transmitted spectrum and in
particular whether a high ionization parameter can be detected at
intermediate stages. In \S~7, we consider the different ionization
time scales of the various elements and the possible effects that this
can have on the estimate of metallicity derived from the fit of the
relative opacities. Finally, in \S~8, we summarize and discuss our
results.

\begin{figure}
\psfig{file=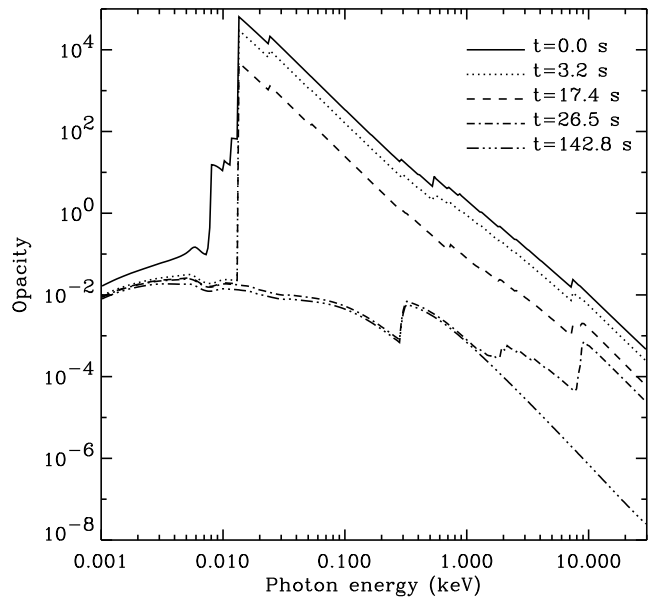,width=0.48\textwidth}
\caption{{The opacity as a function of frequency from optical to 
X-rays in several time shots after the ionizing continuum onset.  The
absorbing medium is a spherical cloud with radius $R=10^{18}$~cm and
uniform density $n=10^4$~cm$^{-3}$ ($N_H=10^{22}$~cm$^{-2}$) with the
source at its centre.  Abundances are set to the solar value and the
dust-to-gas mass fraction is 0.01 (the average Galactic value).}
\label{fig:sim1}}
\end{figure}

\section{Code outputs}

The code we used for this analysis is fully described in Paper I,
where the physics of dust evaporation is discussed in detail. We
consider two species of grains: graphite, purely composed of carbon
atoms, and silicates, made of MgFeSiO$_4$ molecules\footnote{As
discussed in Paper I, the silicate grain should be rather thought as a
lattice with average elemental abundances described by this formula.}.
The effects of dust opacity and destruction were implemented into a
widely used photoionization code (Raymond 1979; Perna, Raymond \& Loeb
2000).

\begin{figure}
\psfig{file=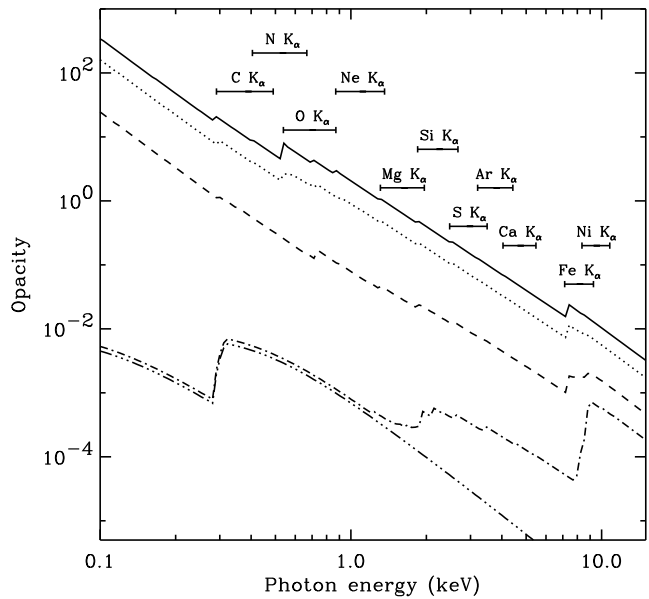,width=0.48\textwidth}
\caption{{Zoom of Fig~\ref{fig:sim1} in the X-ray ([0.1--15]~kev) range.
Horizontal bars show the range of $K_\alpha$ photoionization edge
frequencies for the considered elements (besides H and He) from
neutral to H-like.}
\label{fig:sim2}}
\end{figure}

Among the various outputs of the code described in Paper I, we will
mainly use in this paper the radially integrated opacity as a function
of frequency, which we will call ``spectral opacity distribution''
(SOD). An example output of the code is shown in Figs.~\ref{fig:sim1}
and \ref{fig:sim2}, where the opacity is plotted versus frequency for
a uniform cloud\footnote{Note that in all the code runs presented
here the term cloud refers to the uniform cloud of gas that surrounds
the source (which is at its centre) and not to a cloudlet at a
certain distance from the source.} with radius $R=10^{18}$~cm, density
$n=10^4$~cm$^{-3}$, solar metallicity and dust-to-gas ratio $d_k=1$,
where $d_k$ is the ratio between the mass of dust and the mass of gas,
normalized to the local galactic value $d_{k,~\rm{local}}=0.01$. A
cloud with solar dust-to-gas ratio will therefore have $d_k=1$. The
central source sets on at observed time $t=0$ with a constant
[1~eV--100~keV] luminosity of $L=10^{50}$~erg~s$^{-1}$ and a power-law
spectrum $L(\nu)\propto\nu^0$, which correspond to a photon
luminosity of $Q_H\approx7.2\times10^{57}$~s$^{-1}$. This choice of
luminosity and spectrum is appropriate if the central source is a
Gamma-Ray Burst. This code can be applied in principle to different
objects, such as AGNs, especially to describe the effect of periods of
enhanced activity on the surrounding environment. A rather smaller
luminosity $L\sim~10^{40-45}$~erg~s$^{-1}$ should be adopted in this
case, and we will discuss below (see \S~3.1) in which cases the
results presented here can be simply rescaled to the case of AGNs (see
also Paper II).

\begin{figure}
\psfig{file=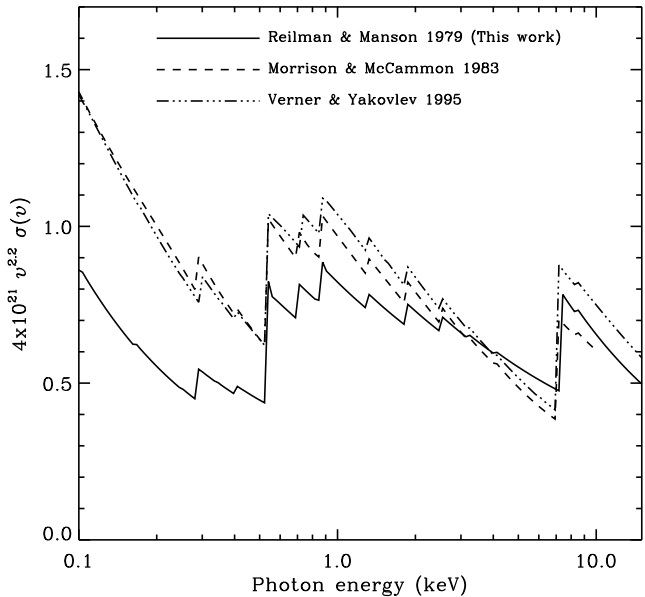,width=0.48\textwidth}
\caption{{Comparison of the cross sections  used in the 
code (Reilman \& Manson 1979) with two other widely used atomic
datasets (Morrison \& McCammon 1983 and Verner \& Yakovlev
1995). Differences, up to a factor 1.5, are present especially in the
range where the absorption is dominated by Helium. Much better
agreement is obtained at larger frequencies. }
\label{fig:comp}}
\end{figure}

In Fig.~\ref{fig:comp} we show a comparison between the SOD of a cold
gaseous cloud as implemented in our code (based on Reilman \& Manson
1979) and two more recent atomic datasets (Morrison \& McCammon 1983
and Verner \& Yakovlev 1995). Here and in the following we will
use indifferently the term {\it cold} or {\it neutral} that will have
the same meaning of non ionized. The agreement is reasonable, even
though some discrepancy in the very soft X-ray range can be seen. This
is mainly due to a different shape for the H and He photoionization
cross-sections at frequencies much larger than the threshold.

\section{The effect of dust on the photoionization of metals}

As mentioned in the introduction, the presence of dust has mainly two
effects on the photoionization rate of the ISM. The first effect is
that of reducing the transmitted UV flux. This causes a delay in the
photo-ejection of higher orbital electrons which is of relatively
small importance. In fact, in the X-ray regime, the opacity is mainly
due to $K$-shell electrons that are not sensitive to UV radiation.

A much more important effect is that of shielding that the grain has
on the metals contained inside the grain itself. This is not due to
the fact that the grain is opaque to X-ray radiation (a condition that
is satisfied only in the very large grains that do not have a sizable
contribution on the absorption) but to the capacity of the grain
itself to sustain an electric potential large enough to avoid the
ejection from the grain of photoelectrons that are therefore
re-captured by the atom that was stripped. In order to understand when
and why this happens it's worth reviewing the grain destruction
mechanisms described in Paper I, to which we remind the reader for a
more thorough analysis.

We consider dust destruction due to two mechanisms: thermal
sublimation and ion field emission (IFE). The first mechanism is
active when the flux of UV and soft X-ray photons is large enough to
heat the grain above a critical temperature of the order of several
thousand Kelvin. The condition is realized if either the grain is
close to the photon source or the spectrum is soft (see \S~4). In
these conditions the grain evaporates by ejecting neutral or singly
ionized atoms in a time scale that is much smaller than the ionization
time scale for the inner electrons. For this reason the X-ray opacity
depends only weakly on whether the atom was initially bound to a grain
or in gaseous phase.

\begin{figure}
\psfig{file=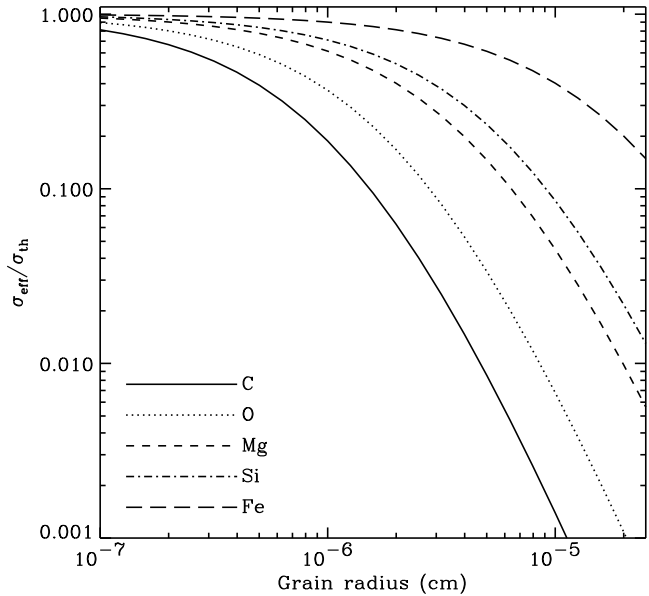,width=0.48\textwidth}
\caption{{Effective cross section for $K_\alpha$ photoionization of elements 
in maximally charged grains vs. the grain radius. The effective cross
section is shown divided by the cross section of the same atom in
gaseous phase, in order to emphasize the effect of metals contained in
grains versus those in gaseous phase.}
\label{fig:sef}}
\end{figure}

If the grain temperature is not kept large enough, the thermal
sublimation time scale increases exponentially and a different
mechanism (IFE) can set in. In particular, if the ionizing continuum
has a hard spectrum, intermediate-hard X-ray photons can ionize the
grain without sizably heating it. Due to the strength of the molecular
bonds, the charged grain can resist up to a threshold potential of
$3\,a_{-5}$~kV (Waxman \& Draine 2000) where $a_{-5}$ is the grain
radius in units of $10^{-5}$~cm. After this threshold is
overcome, the grain reacts to any further ionization by ejecting an
ion. Consider, however, a photon with $h\nu=0.3$ keV and a graphite
grain. The photon has energy large enough to photoionize an isolated
carbon atom by stripping a $K$ electron that has an ionization energy
of 0.291~keV. If however the carbon atom is bound in a charged
graphite grain, the photoelectron will be unable to escape from the
grain due to the large potential barrier and will consequently
recombine in a very short time with a carbon ion. A carbon atom bound
to a maximally charged graphite grain of radius $a$ has therefore an
effective ionization energy:
\begin{equation}
h\nu_{\rm{eff}}=h\nu_{\rm{th}}+3\,a_{-5} \,\,{\rm keV}
\end{equation}
where $h\nu_{\rm{th}}$ is the ionization potential of the isolated
atom. This will correspond also to a decrease in cross section:
\begin{equation}
{{\sigma_{\rm{eff}}}\over{\sigma_{\rm{th}}}} \sim 
\left(1+{{3\,a_{-5}\,{\rm{kev}}}\over{h\nu_{\rm{th}}}}\right)^{-3}
\label{eq:sef}
\end{equation}
where $\sigma_{\rm{th}}$ is the threshold cross section for
photoionization of an isolated atom. The decrease in the cross
section, which is proportional to the increase in photoionization
time scale (Lazzati, Perna \& Ghisellini 2001), is shown in
Fig.~\ref{fig:sef} for C, O, Mg, Si and Fe, the atoms that we consider
as components of graphite and silicates. In the figure, the proper
shape of the cross section function is taken into account, rather than
the simplified power-law behavior of Eq.~\ref{eq:sef}. It is clear
that, the lower the photoionization potential, the larger is the
decrease of the effective cross section, and therefore graphite grains
will be more effective in shielding the photoionization with respect
to silicates, in which iron is almost unaffected.

Another effect in competition with IFE grain destruction is the
recombination of free electrons onto the charged grain, when its
charge is below the threshold for the IFE effect (Fruchter et al. 2001;
Paper I). Even if the importance of recombination with respect to
ionization depends on the type of grain considered and on the spectrum
of the ionizing source, one can roughly write the ratio of the
time scales as:
\begin{equation}
{{dN_{\rm{ion}}/dt}\over{dN_{\rm{rec}}/dt}}\sim 10^{-8} \, \xi
\, T_5^{1/2}
\end{equation}
where $\xi\equiv{}L/(n\,R^2)$ is the ionization parameter (see
also~\S~6) and $T_5$ is the electron temperature in units of
$10^5$~K. The effect of recombination can be therefore neglected in
our computations (it is marginally important for the largest and
densest could, see \S~6), but should be taken into account when
considering less luminous ionizing sources and/or absorbers at larger
distances.

\begin{figure}
\psfig{file=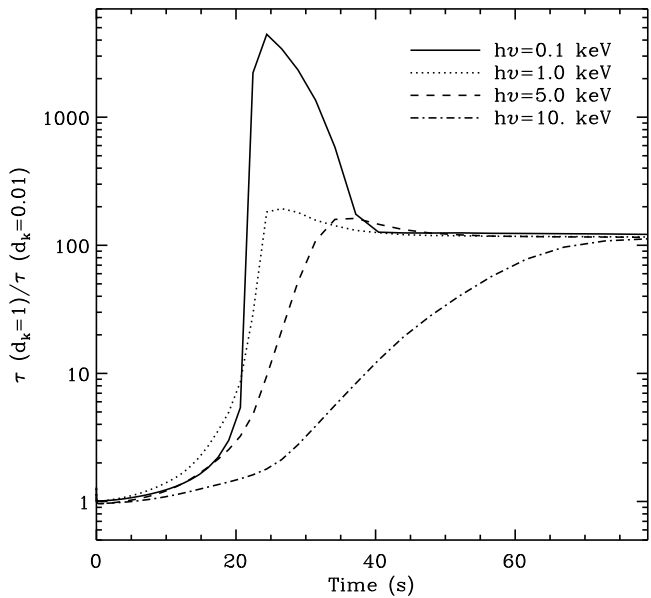,width=0.48\textwidth}
\caption{{The effect of dust in the evolution of the X-ray opacities. 
The figure shows a comparison between the opacity at various
frequencies for a simulation with standard dust-to-gas ratio and a
simulation with a hundred times less dust. In both the simulations the
absorbing medium is a spherical cloud with radius $R=10^{18}$~cm and
uniform density $n=10^4$~cm$^{-3}$ ($N_H=10^{22}$~cm$^{-2}$).
Abundances are set to the solar value. }
\label{fig:dnd}}
\end{figure}

The effect of dust in changing the ionization time scales is shown in
Fig.~\ref{fig:dnd}, where the opacities at selected frequencies versus
time are compared for two sets of simulations: one with solar
dust-to-gas ratio ($d_k=1$) and the other with a dust depletion of a
factor 100 ($d_k=0.01$).  The opacity in the latter, ``no-dust''
simulation decreases on a smaller time scale, and in fact after
$\sim20$ seconds, the opacity in the dusty case  can be more than
on thousand times larger than in the no-dust case (see the
$0.1$~keV line in Fig.~\ref{fig:dnd}). At very late times, the SOD in
both simulations is dominated by opacity due to unevaporated graphite
grains, and therefore in the no-dust simulation is 100 times smaller
than in the dusty one (see however \S 4 for how the effect is reduced
with a softer spectrum).  The effect can also be seen in
Figs.~\ref{fig:sim1} and~\ref{fig:sim2}: the lower line, which shows
the opacity at the largest time, is completely dominated by the
opacity due to the large graphite grains, which survived destruction
at large radii by the shielding mechanism.  Should such an extreme
condition be realized in a real source, it may be recognizable, being
characterized by a typical $N_H/A_V$ ratio (see also Paper I and II)
\begin{equation}
\left({{N_H}\over{A_V}}\right)_{\rm{graphite}} \sim 10^{20}\,{\rm{cm}}^{-2}
\end{equation}
which is roughly ten times smaller than for a cold gas with solar
metallicity and dust content.  Moreover, it should be characterized by
a prominent\footnote{The edge may become less and less important for
very big grains of radius larger than $\sim1\mu$m.} absorption edge at
the threshold of neutral carbon: $h\nu=0.291$~keV.

Such a phase of the ISM, in which dust grains coexist with a fully
ionized gas, may not be an equilibrium condition. In fact, the
electron temperature of the free electrons will be around
$T\sim10^5$~K (Perna et al. 2000), a temperature large enough to
sublimate the grains. The energy input rate of the grains due to the
electron bombardment is ${\dot E} = \pi\,a^2\,n_e\,v_e\,\epsilon_e$
where $v_e$ is the electron mean velocity and $\epsilon_e$ their mean
energy. This heating term is always much smaller than the black-body
cooling term ${\dot E}=4\pi\,a^2\,\sigma\,T^4$. The grain is therefore
stable in the hot electron bath, at least for the time scale of
interest in this paper.

\begin{table*}
\begin{tabular}{cl}
Symbol & Explanation \\ \hline \hline $N_X$ & Total column of the
element $X$, irrespective of its ionization state, even if completely
stripped off all the electrons. \\ $N_{X>}$ & Total photo-absorbing
column of the element $X$, i.e. with at least one associated
electron. \\ $N_{[X]}$ & Column density as measured from soft X-ray
absorption, under the assumption of a cold intervening medium.
\\ 
$R$ & Radius of the surrounding cloud \\
$n$ & Density of the absorbing medium \\
$L$ &Luminosity of the power-law ionizing continuum in the 
[1~eV--100~keV] range \\
$\alpha$ & Spectral index of the power law continuum 
($L(\nu)\propto\nu^{-\alpha}$) \\
$a$ & Radius of dust particles \\
$\xi$ & Ionization parameter of the gas \\
$d_k$ & Dust content parameter: ratio of dust-to-gas 
mass densities relative to the solar value
\end{tabular}
\caption{{Explanation of symbols used throughout the text.}
\label{tab}}
\end{table*}

\begin{figure}
\psfig{file=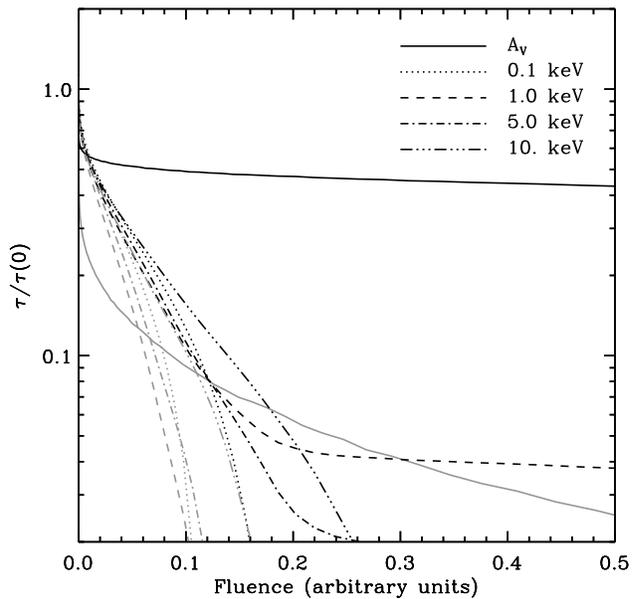,width=0.48\textwidth}
\caption{{Opacity as a function of fluence for a cloud surrounding a
uniform source of ionizing photons (dark lines) and a source with 10
times larger peak luminosity but which is on only for $10\%$ of the
time, so that the two sources have the same average fluence ${\cal
F}(t)$.  See \S 3.1 for more details.}
\label{fig:var}}
\end{figure}

\subsection{Breaking the time, size and luminosity degeneracy and the 
effect of a variable source}

\begin{figure*}
\parbox{0.49\textwidth}{
\psfig{file=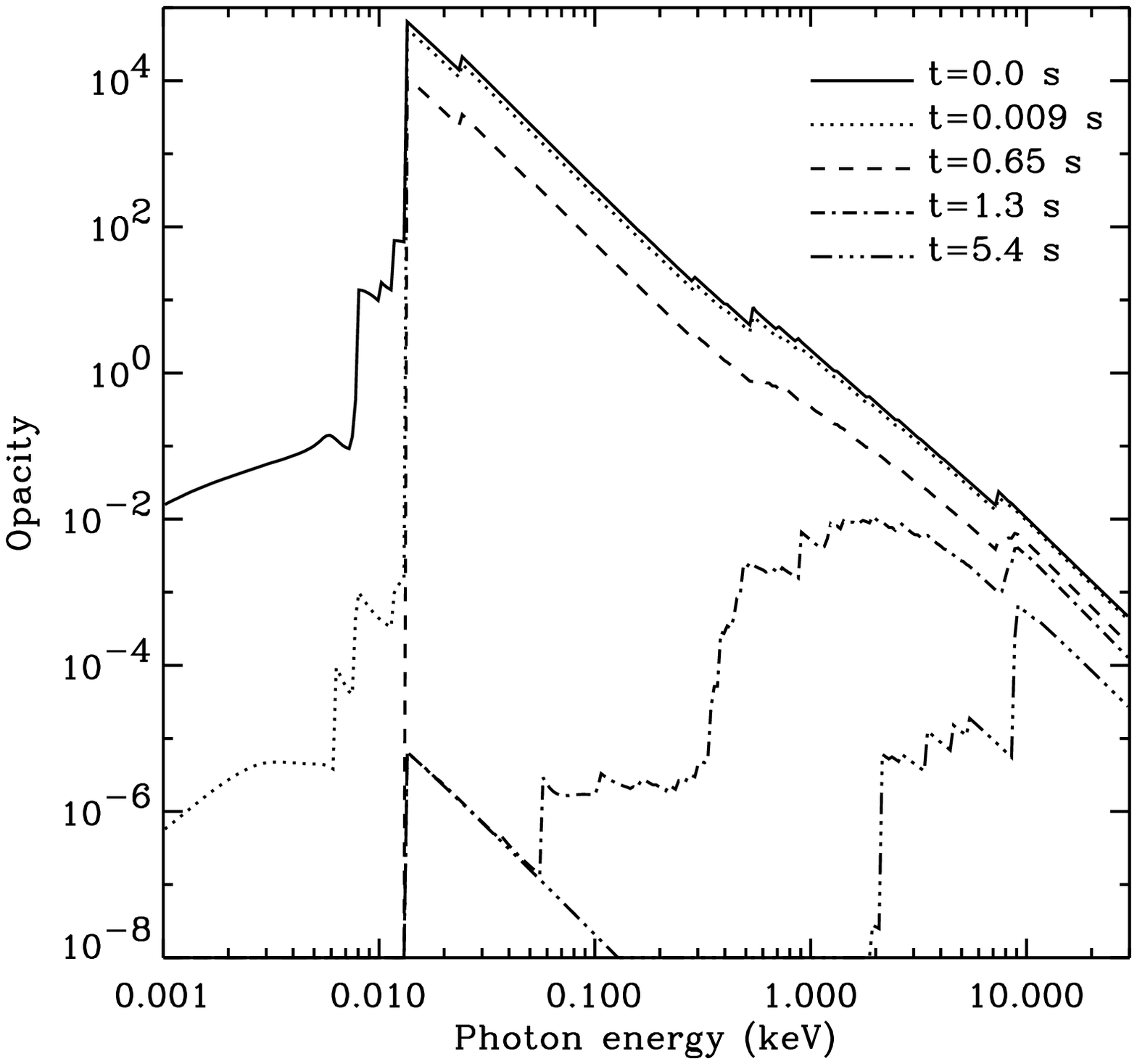,width=0.48\textwidth}
\caption{{Same as Fig~\ref{fig:sim1} but for $\alpha=0.5$.}
\label{fig:sima}}}
\parbox{0.49\textwidth}{
\psfig{file=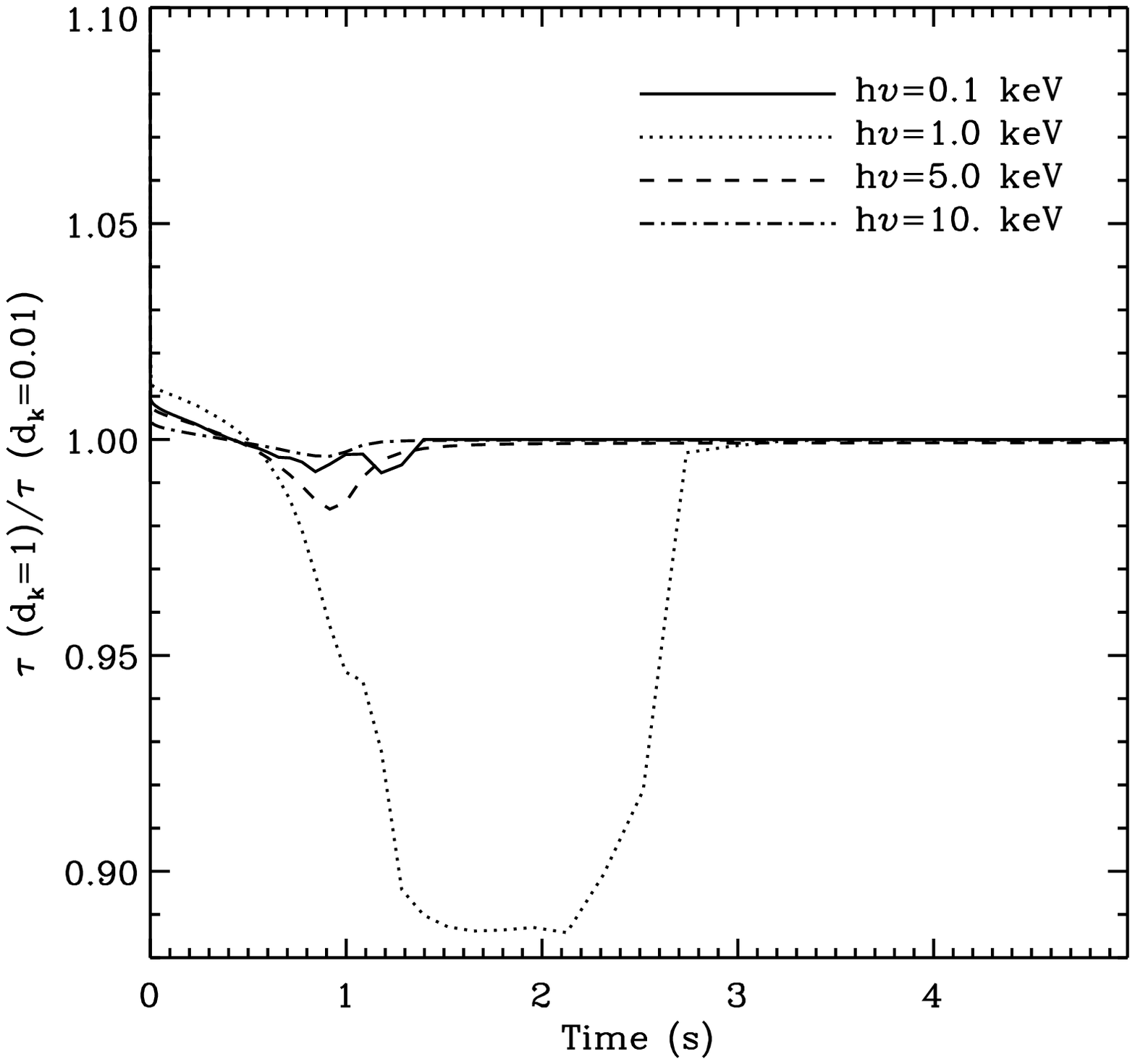,width=0.48\textwidth}
\caption{{Same as Fig~\ref{fig:dnd} but for $\alpha=0.5$.}
\label{fig:dnda}}}
\end{figure*}

When a purely gaseous cloud is considered, it can be shown that the
ionization state of the cloud depends on the local fluence rather than
on the flux and time elapsed since the source onset independently
(LP02). This is an important conclusion, which holds in both the
optically thin and thick cases, since it allows to apply the results
of a simulation with a given parameter set to a different cloud
geometry and input luminosity (see Eq. 1 in LP02). This relation also
implies that a simulation with constant flux can be adapted for a
variable source when the time is properly defined (see discussion in
LP02). Moreover, a simulation with a given luminosity can be adapted
to any other luminosity by simply rescaling the time, provided that it
does not become longer than the recombination time scale.

When dust grains are introduced as a component of the absorbing
medium, the degeneracy is broken and these scaling laws are no more
valid. This is due to the fact that the degree of destruction of a
dust particle does not depend uniquely on the number of photons that
it absorbed, but also on the rate with which the photons have been
absorbed. This is due to the fact that dust particles can radiatively
cool and have a well defined temperature, the mean value of which
is set by a balancing of the heating and cooling rates.  For this
reason, a highly variable source will be much more effective in
destroying dust than a uniform one with the same total energy and
overall duration.

The bottom line of this is that two luminosity and geometrical setup
are now equivalent if i) the column density is the same and ii) the
local flux is the same, i.e. $L_1/R_1^2 = L_2/R_2^2$, where $L_1$ and
$L_2$ are the luminosity of the central sources and $R_1$ and $R_2$
the cloud radii in the first and second simulation, respectively. In
other words, the results shown in Fig.~\ref{fig:sim1} can be applied to an
AGN with the same spectrum but luminosity $L=10^{44}$~erg~s$^{-1}$
surrounded by a cloud of radius $R=10^{15}$~cm and density
$n=10^7$~cm$^{-3}$.

This holds for a uniform luminosity, while additional complications
are required in the case of a variable source. In Fig.~\ref{fig:var}
we show the effect of variability in the ionizing luminosity for
our standard cloud with radius $R=10^{18}$~cm, density
$n=10^4$~cm$^{-3}$ and standard dust-to-gas ratio $d_k=1$. The central
source has, in both cases, a power-law spectrum
$L(\nu)\propto\nu^0$. With black lines we show the opacity as a
function of fluence, for selected frequencies, for a source that turns
on at $t=0$ and then radiates uniformly in time with a luminosity
$L=10^{50}$~erg~s$^{-1}$. With gray lines, we show the opacity as a
function of fluence for a source that has a peak luminosity ten times
larger, but is intermittent after the onset, being active only for $10\%$
of the time. The two sources have, on average, the same fluence at
time t, so that the x-axis of the figure is proportional to the elapsed
time for both sources. For this reason, the black and gray lines would
overlap for a purely gaseous cloud. In the variable source case,
however, dust grains are held at a much larger temperature, so that
they can evaporate much faster. This is clearly visible by comparing
the solid lines in Fig.~\ref{fig:var}, which show the absorption in
the optical $V$ band, largely dominated by dust: it is asymptotically
more than ten times smaller for the variable source than for the
steady one.

\begin{figure}
\psfig{file=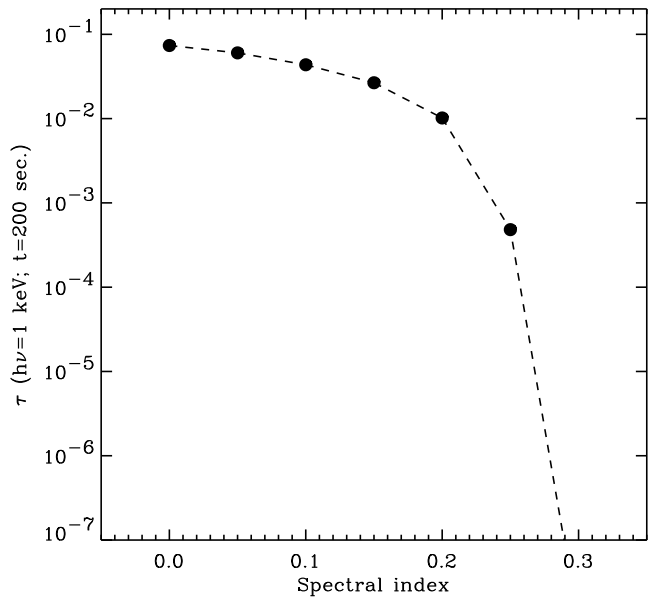,width=0.48\textwidth}
\caption{{Late time opacity at 1~keV for a set of simulations
with different spectral slopes of the ionizing continuum. The
absorbing cloud is always characterized by $R=10^{18}$~cm,
$n=10^4$~cm$^{-3}$ and $d_k=1$. The opacity at 1~keV at $t=200$
seconds is dominated by residual graphite grains in all simulations
with $\alpha\lsim0.25$, while vanishes for softer spectra. See text
for more details.}
\label{fig:alf}}
\end{figure}

\section{Dependence on spectral index}

The ionization state of the absorbing medium and the effectiveness of
an ionizing flux in destroying dust grains depend strongly on
the spectral energy distribution (SED) of the ionizing continuum. A
softer spectrum will be more effective in destroying dust through
thermal sublimation, while a hard spectrum will photoionize metals
with good efficiency, but will leave most of the dust grains
unaffected. In both cases, the SOD will be at the end very different
from the initial one.

A very soft spectrum will leave behind a dust clean environment, with
opacity largely suppressed at all frequencies but for the keV band, in
which residual absorption from iron may still be present. On the other
hand, a very hard spectrum will initially suppress the high frequency
opacity until only dust grains are left which resist to further
destruction thanks to the potential barrier discussed above. The
different behaviors can be seen by comparing the figures obtained with
different spectra. We have produced figures for power-law incident
spectra $L(\nu)\propto\nu^{-\alpha}$, with luminosity
$L=10^{50}$~erg~s$^{-1}$ in the [1~eV--100~keV] band. The spectral
index was either $\alpha=0$, which should reproduce the low frequency
SED of a typical GRB or $\alpha=0.5$, which should be more suited for
a GRB with an optical flash\footnote{ In this case the photon
luminosity would be $Q_H\approx2\times10^{59}$~s$^{-1}$.} (see also
Draine \& Hao 2002).  Fig.~\ref{fig:sim1}, \ref{fig:sim2} and
\ref{fig:dnd} have been produced with the harder spectrum, while
Fig.~\ref{fig:sima} and
\ref{fig:dnda} are relative to the softer one.

Of particular interest is the comparison of Fig~\ref{fig:sim1} with
Fig~\ref{fig:sima}. Both figures show the SOD of the same cloud in
several time shots, but the former is ionized by a hard spectrum,
while the latter is ionized by a softer one. Note that the times at
which the SOD are shown are different, since the soft spectrum has a
larger number of photons and is therefore a more efficient source of
ionization. The dotted line, in both figures, shows the evolution of
the SOD at early times, when the absorption properties of the cloud
start to be modified. Comparing the dotted lines in
Figs.~\ref{fig:sim1} and~\ref{fig:sima} above and below the Lyman
limit, it is easy to note how the soft spectrum is effective in
evaporating dust particles. In comparison, a harder spectrum is much
less effective, so that at late times the opacity is dominated by dust
grains (dash-dotted lines in Fig.~\ref{fig:sim1}) while in
Fig.~\ref{fig:sima} only a residual iron opacity is left at late
times.

The hardness of the spectrum is also an extremely important parameter
in evaluating to which extent the presence of dust influences the
evolution of the X-ray opacity (comparison of Figs.~\ref{fig:dnd}
and~\ref{fig:dnda}). In fact, a soft spectrum efficiently destroys
dust grains before the ionization state of the atoms in the gaseous
phase is modified and therefore the ionization of elements takes place
in a gas free environment even if it started with a sizable amount of
dust. On the other hand, if the spectrum is not soft enough to destroy
dust grains, the presence of dust (which is initially virtually
undetectable from an X-ray spectrum) has a tremendous importance in
determining the evolution of the X-ray SOD (see Fig.~\ref{fig:dnd}).
The key parameter is therefore the ratio of the photoionization time
scale to the grain destruction time scale which is a function of the
spectral index of the spectrum and of the distance from the source (or
the source luminosity).

It is also important to note that the amount of dust that is leftover
for a cloud with a given geometry and density and for a certain
ionizing luminosity is a threshold effect. Either the ionizing flux is
capable of destroying all the dust grains or most of them are left
behind, the intermediate situation being possible only in a very
limited range of spectral shapes. In Fig~\ref{fig:alf} we show the
opacity at 1~keV at the largest simulated time ($t=200$~s) as a
function of the spectral index $\alpha$ in a set of simulations for
which only $\alpha$ is varied. The geometry of the absorbing cloud is
given by $R=10^{18}$~cm, $n=10^4$~cm$^{-3}$ and $d_k=1$ while the
central source has a luminosity $L=10^{50}$~erg~s$^{-1}$. The choice
of plotting the opacity at this frequency is due to the fact that at
late times (in this case 200 seconds) the opacity at 1~keV (if any) is
dominated by absorption by dust particles (mostly big graphite grains,
see Fig.~\ref{fig:sim1}). For spectra softer than $\alpha=0.25$, all
the grains are destroyed and there is no residual opacity, while for
harder spectra a considerable fraction of grains is left behind.

\begin{figure*}
\parbox{0.33\textwidth}{\psfig{file=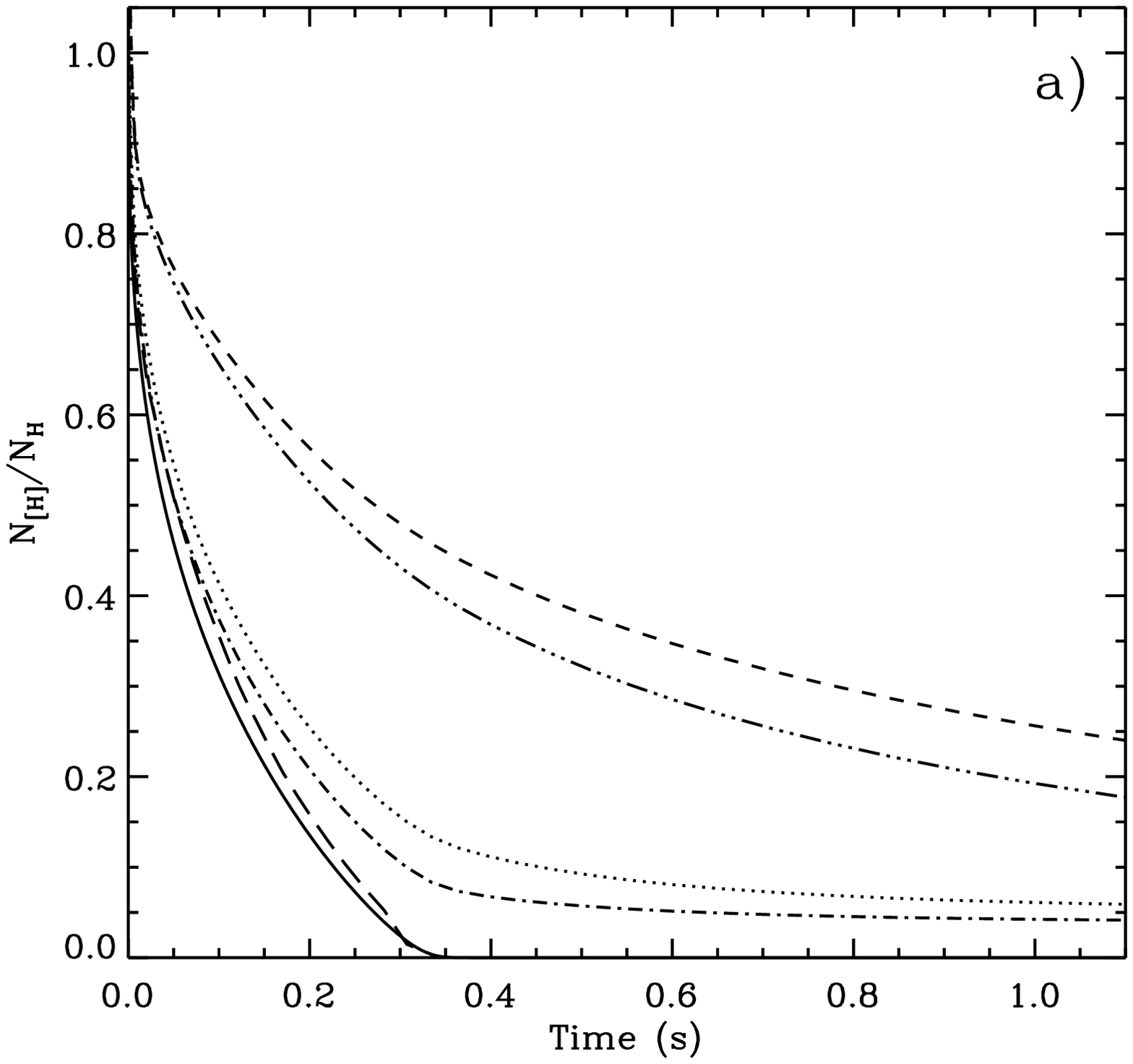,width=0.32\textwidth}}
\parbox{0.33\textwidth}{\psfig{file=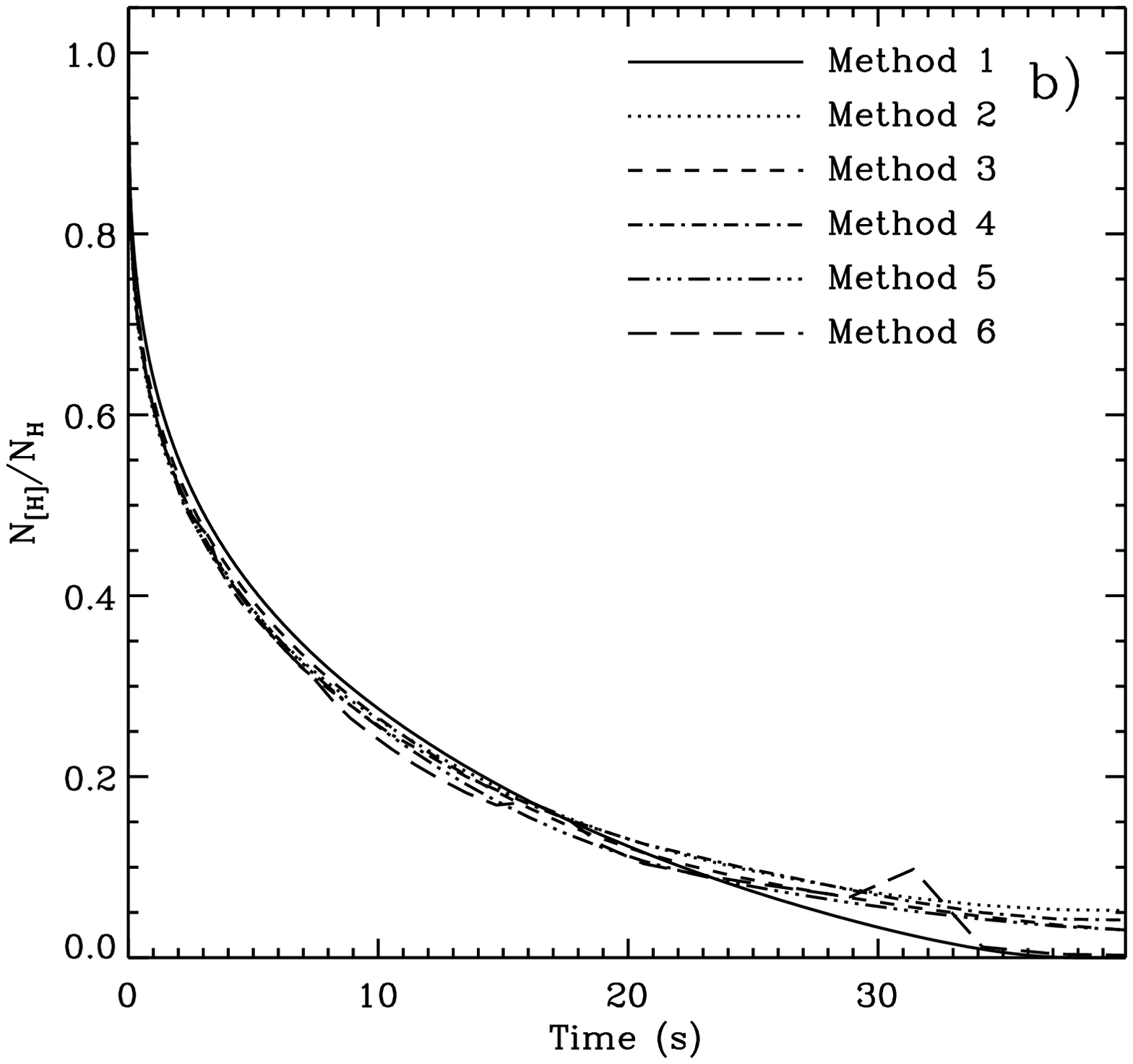,width=0.32\textwidth}}
\parbox{0.33\textwidth}{\psfig{file=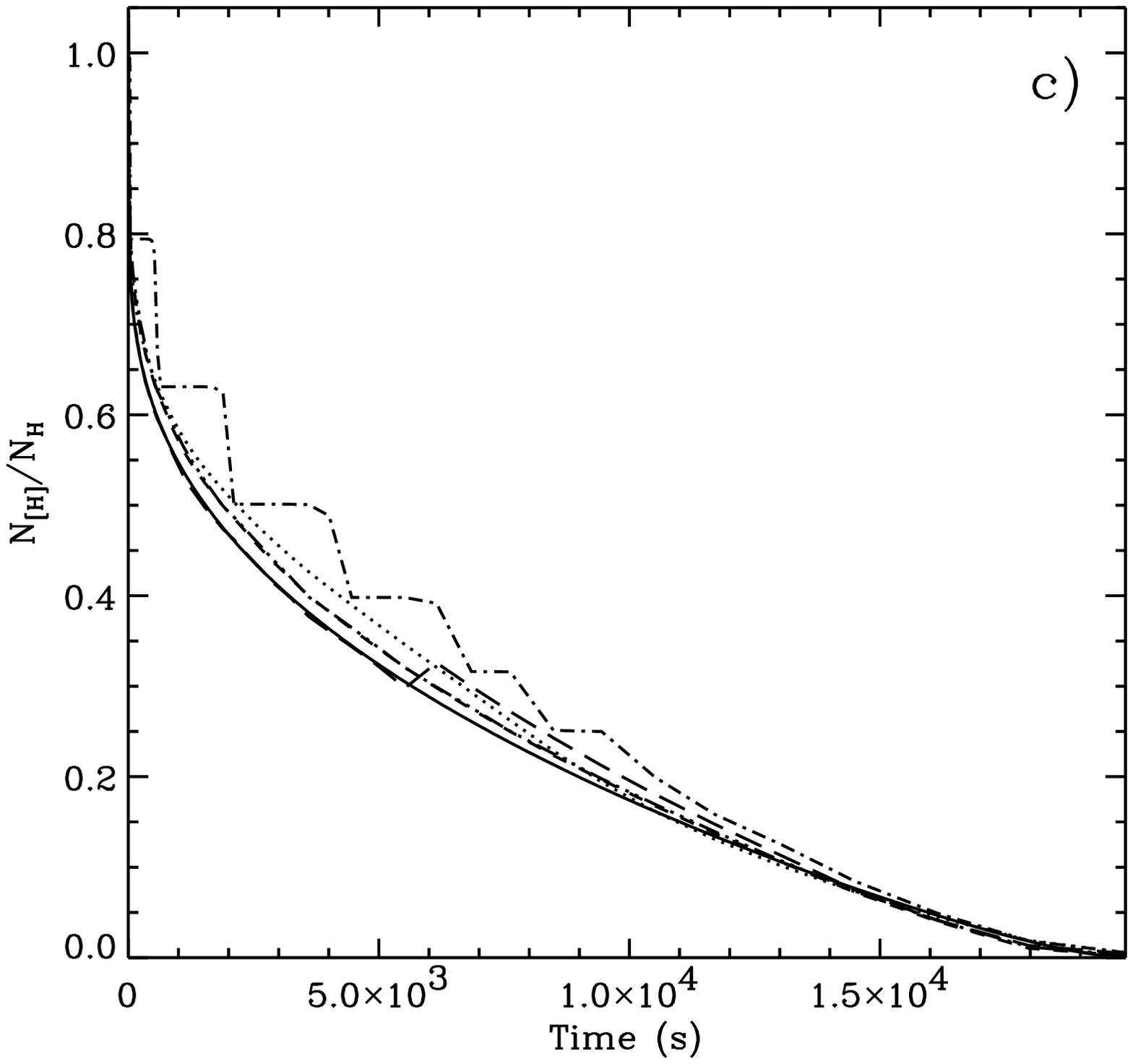,width=0.32\textwidth}}
\caption{{Comparison of the 6 methods for the evaluation of the 
effective $N_H$ for various initial column densities. Results shown
here are for a cloud with $R=10^{18}$~cm, dust-to-gas ratio $d_k=1.$
and for a central source with $L=10^{50}$~erg~s$^{-1}$ and
$\alpha=0$. Panel a), b) and c) show initial column densities
$N_H=10^{20}$, $10^{22}$ and $10^{24}$~cm$^{-2}$, respectively.}
\label{fig:nhs}}
\parbox{0.33\textwidth}{\psfig{file=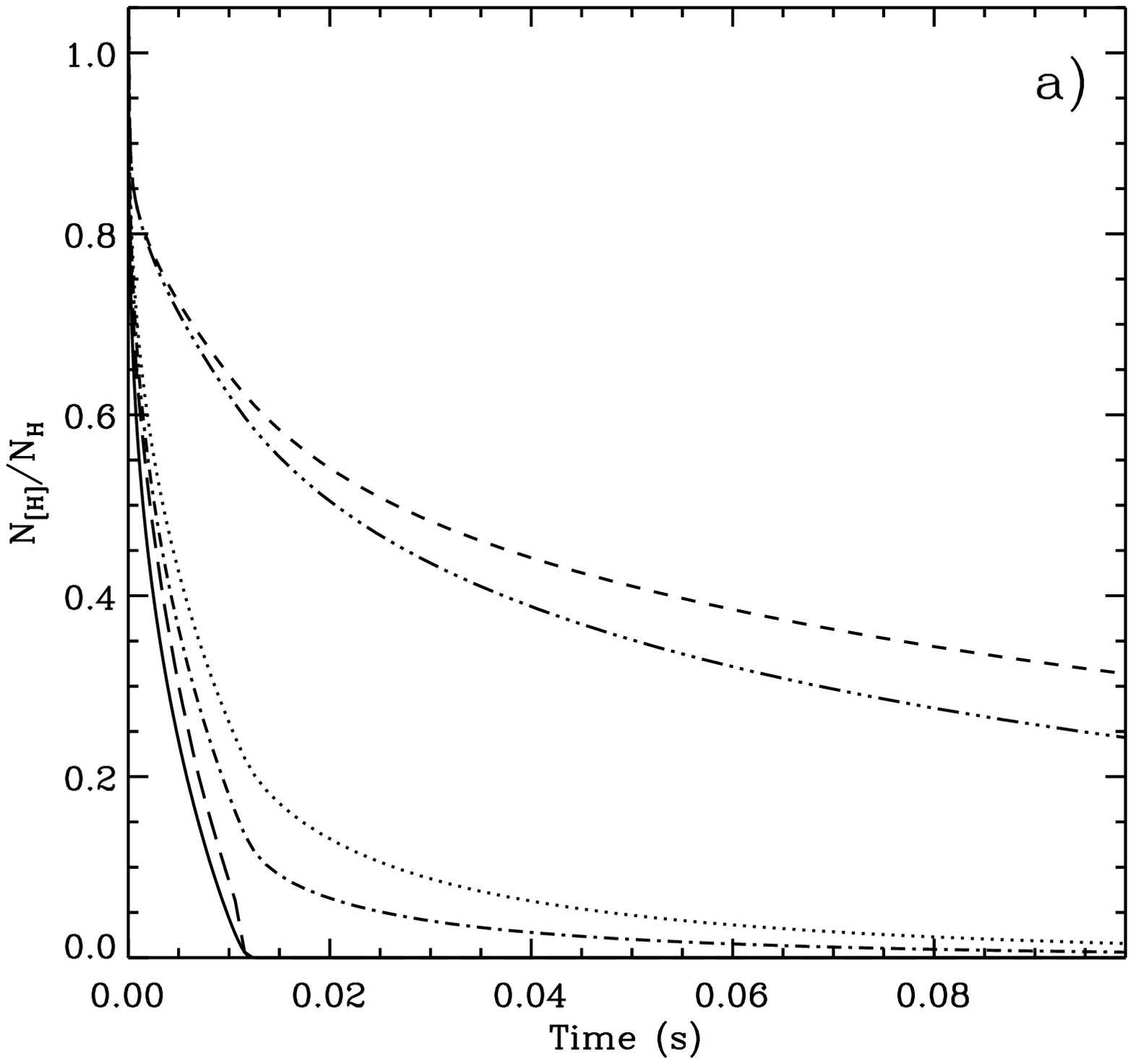,width=0.32\textwidth}}
\parbox{0.33\textwidth}{\psfig{file=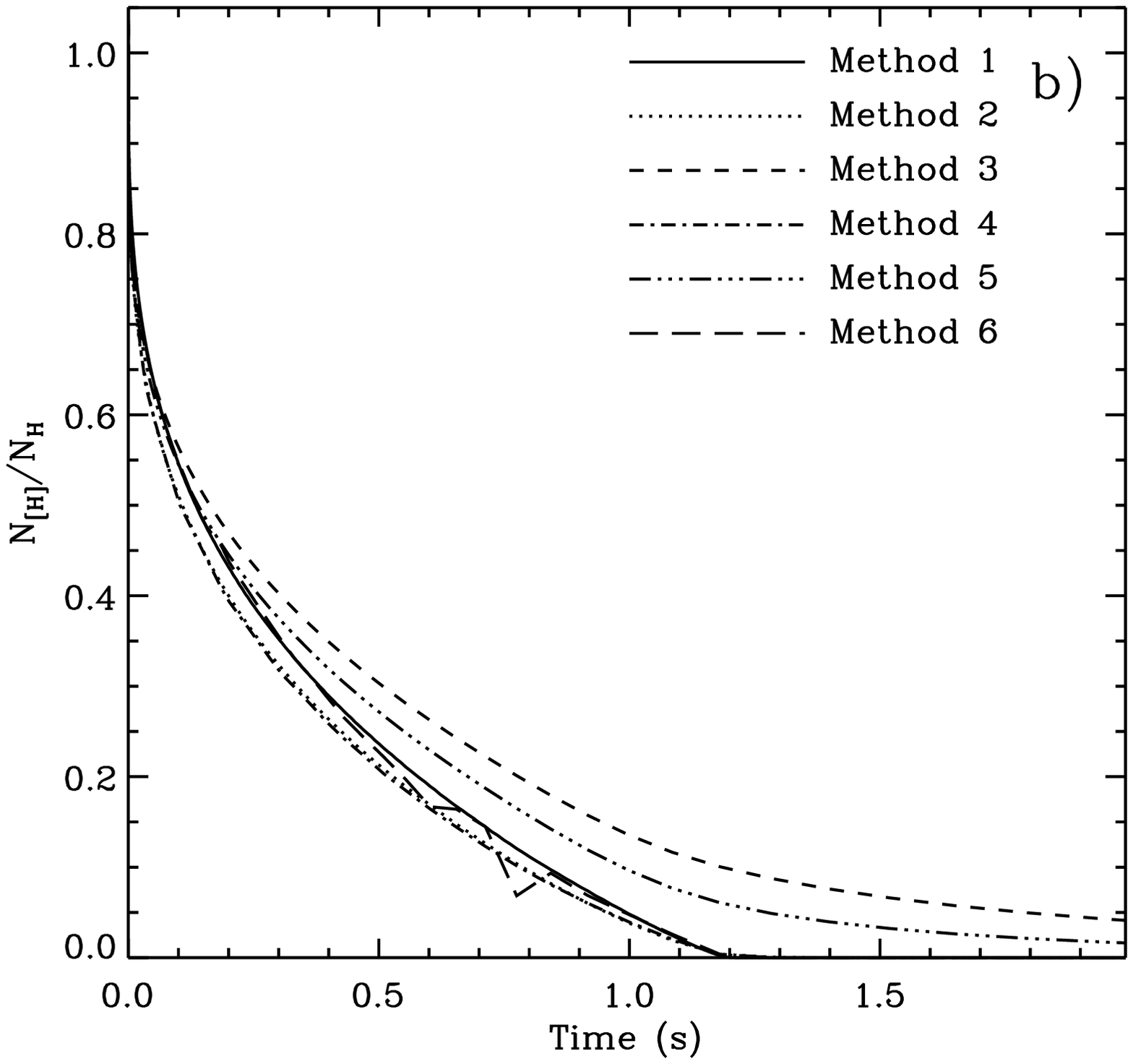,width=0.32\textwidth}}
\parbox{0.33\textwidth}{\psfig{file=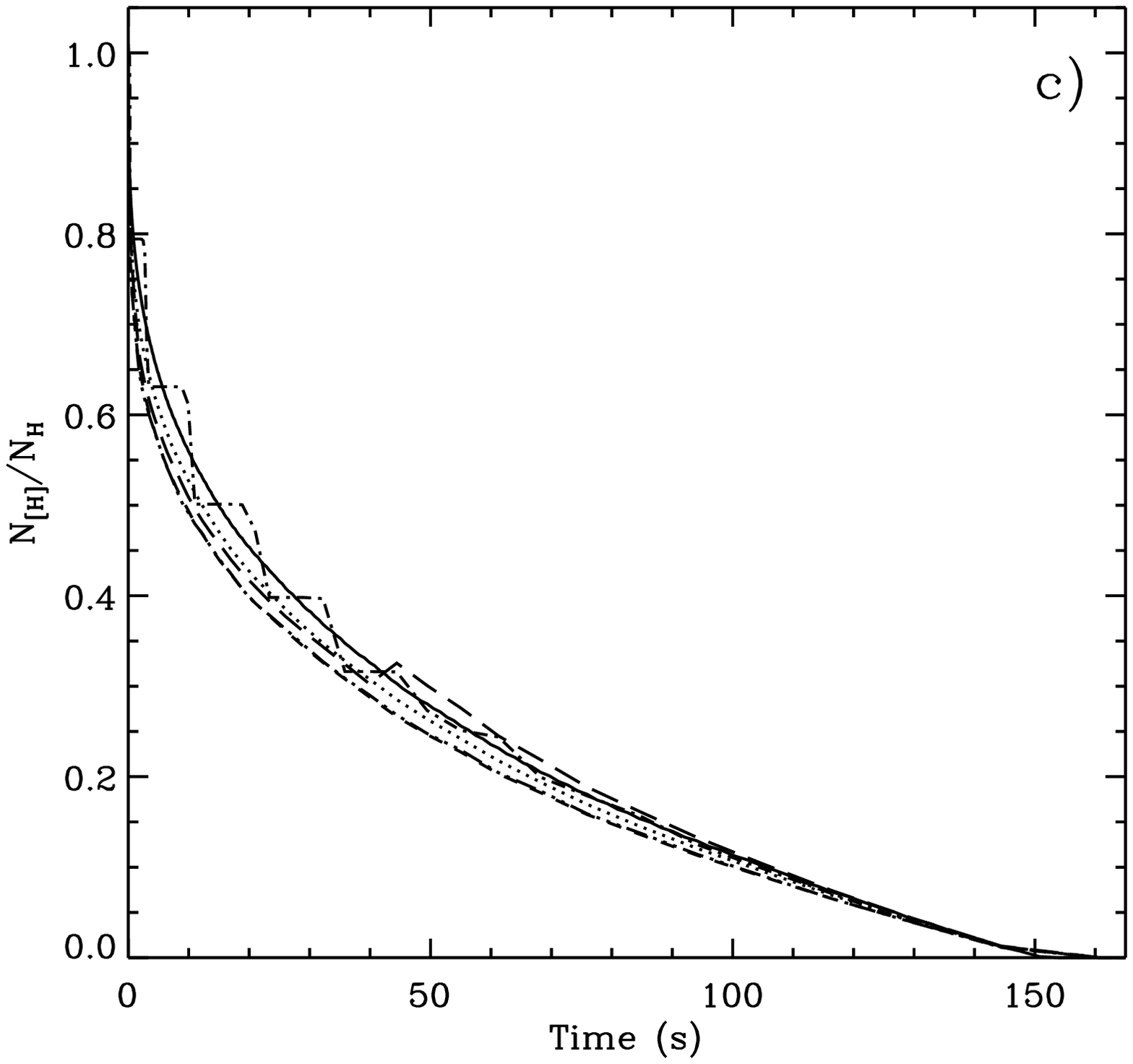,width=0.32\textwidth}}
\caption{{Same as Fig.~\ref{fig:nhs} but for $\alpha=0.5$. }
\label{fig:nhsa}}
\end{figure*}

\section{The measure of the residual column density}

As discussed in LP02, it is important to convert the result of the
simulations, which contain very detailed informations about the
evolution of the frequency dependent opacity with time, in a single
indicator to be compared to observed spectra, especially if the signal
to noise of them does not allow for the comparison with a detailed
model. It is customary in the X-rays to quantify the soft X-ray
absorption through the amount of column density of hydrogen $N_{[H]}$,
under the assumption that the intervening medium is cold (see, e.g.,
Morrison \& McCammon 1983).

The hydrogen column density is a good measure for the amount of
absorption if the metallicity of the ISM is known and if the ISM is
neutral. In fact, the measure of the column density is made through
the comparison of the observed spectrum with an average
photoabsorption cross section, computed under the assumption of solar
metallicity for neutral elements. The assumption of a completely
recombined gas is clearly not adequate in our case. We have developed
several methods to derive an ``effective column density'' from the
simulated spectra, which we will discuss and compare in detail
below. The output of the various methods are also compared in
Figs.~\ref{fig:nhs} and~\ref{fig:nhsa} for several cloud geometries
and two spectral indices of the ionizing continuum. Before doing that,
it is however worth defining some quantities which are relevant when
part of the atoms are completely stripped of their electrons and do
not contribute to the absorption any more. We will call in the
following $N_X$ the column density of the element $X$ irrespective of
its ionization state, even if the element has lost all its electrons
and does not absorb photons any more. These quantities are a property
of the cloud that is not modified by the central ionizing source. To
indicate the quantity of absorbing ions of the element $X$, we use the
notation $N_{X>}$, which is the column density of atoms and ions of
the element $X$ with at least one electron. Finally, we will use the
symbol $N_{[H]}$ for the effective column density, i.e. the equivalent
column of cold hydrogen that produces an amount of absorption
comparable to what is obtained in the simulation. The following of
this section is dedicated to find a good way to compute this quantity.

\subsection{Real column density of absorbing hydrogen ($N_{H>}$)}

As an output of our simulations, it is possible to derive the
absorbing column density of HI as a function of time ($N_{H>}$). This
is not really a prescription for comparing the observations with the
simulations. Rather, the comparison of this $N_{[H]}$ with the other
values derived with different methods is an indication of how the
effective metallicity is changed by photoionization. Column densities
derived with this method are shown with solid lines in
Figs.~\ref{fig:nhs} and~\ref{fig:nhsa} (Method 1).

\subsection{Average opacity in selected bands}

Following LP02, the effective column density can be computed by
averaging the value of the opacity in a selected frequency range, and
dividing the obtained value by the average of the cross section of a
cold absorber in the same range:
\begin{equation}
N_{[H]} \equiv \left\langle {{\tau}\over{\sigma_0}} 
\right\rangle_{[\nu_a,\nu_b]}
\label{eq:nh1}
\end{equation}
where $\tau$ is the frequency dependent opacity, $\sigma_0$ is the
(frequency dependent) cross section of a cold medium with solar
metallicity and $\langle~\rangle_{[\nu_a,\nu_b]}$ represents the
average over the frequency range $\nu_a$, $\nu_b$.  The result is
however highly dependent on the frequency range considered, so that
different averages should be compared to measurements performed with
different satellites. As an example we have computed the effective
column densities in the range [0.2--1]~keV (Method 2) and [2--10]~keV
(Method 3). The former is relevant for a typical X-ray instrument with
good sensitivity in the sub-keV range (e.g. the BeppoSAX MECS and
LECS), while the latter is more relevant for instruments such as the
wide-field cameras on board BeppoSAX, whose sensitivity is relevant
above 2 keV. The results of Method 2 and 3 are shown in
Figs.~\ref{fig:nhs} and~\ref{fig:nhsa} with a dotted and a dashed
line, respectively.

\subsection{Average transmitted flux in selected bands}

It can be argued that an instrument is usually not sensitive to the
value of the opacity, but to the transmitted flux. In other words,
experimentally it is difficult, if not at all impossible, to
distinguish between $\tau=5$ and $\tau=500$. For this reason, the
equivalent absorbing column should be derived by averaging in selected
bands the transmitted flux ratio, i.e. $e^{-\tau}$ rather than the
opacity itself. The ``effective column density'' $N_{[H]}$ can
therefore be implicitly defined as:
\begin{equation}
\left\langle e^{-\tau} \right\rangle_{[\nu_a,\nu_b]} =
\left\langle e^{-\sigma_0\,N_{[H]}} \right\rangle_{[\nu_a,\nu_b]}
\end{equation}
where the meaning of symbols is the same as in Eq.~\ref{eq:nh1}.
This method, even though more suited for a comparison with real data,
will return a value of $N_{[H]}$ dependent on the frequency range over
which the average is performed. Again, we performed the calculation in
the [0.2--1]~keV (Method 4) and [2--10]~keV (Method 5) ranges. The
results are shown in Figs.~\ref{fig:nhs} and~\ref{fig:nhsa} with a
dot-dash and a 3-dot-dash line, respectively.

\subsection{Threshold opacity}

An alternative way to estimate an ``effective'' column density that
does not require the definition of a frequency range over which an
average should be taken can be defined. The key of this method is to
consider as more important the frequency range where the opacity is
close to unity, i.e. where its presence is easily detectable and can
be quantified. In fact, if the opacity is too small, there is no
decrement in the transmitted flux while, if the opacity is too large,
no flux is transmitted and only a lower limit can be set.  Since the
cross section for photoionization of a solar metallicity ISM is a
steep function of frequency, the frequency at which the opacity has a
fixed value is well constrained. We therefore computed the frequency
at which the opacity is equal to 0.5 in our simulations and defined
the effective column density $N_{[H]}$ as the inverse of the cross
section of a cold absorber at the same frequency (Method 6). The
results are plotted in Figs.~\ref{fig:nhs} and~\ref{fig:nhsa} with a
long-dashed line. The value 0.5 has been arbitrarily chosen. The
inferred value of $N_{[H]}$ is however only slightly dependent on the
choice of this number, which should be reasonably between 0.1 and 1.
As can be seen from the figures, the disadvantage of this method is
that in some points the value of $N_{[H]}$ has instabilities, related
to the coincidences of the frequency at which the opacity is equal to
0.5 (or whatever selected value) with the photoionization threshold of
an abundant element. In these cases, the cross column density of the
cold absorber for the given opacity is not uniquely defined.

\subsection{Discussion}

The comparison of the different methods described above can be made
from Figs.~\ref{fig:nhs} and~\ref{fig:nhsa}. For the case of an
intermediate and large initial column density
($N_H\ge10^{22}$~cm$^{-2}$), the 6 methods agree quite well, yielding
comparable values of the effective column, irrespective of the
spectral index and, therefore, also of the presence of dust. Method 4
is not particularly suited for large column densities, since in the
low frequency range virtually no flux is transmitted for column
densities larger than $10^{23}$~cm$^{-2}$.  In the case of a
relatively small initial column density, the methods based on averages
at large frequencies (method 3 and 5) tend to give larger column
densities, in particular in the case of a direct average of the
opacity (Method 3). This is due to the fact that heavy metals are
ionized with a slower speed with respect to hydrogen in the optically
thin case. However, in these simulations, the opacity at large
frequencies is very small and therefore methods 3 and 5 are not
relevant since an instrument sensitive in the [2--10]~keV regime would
not detect any absorption.

\section{Fitting spectra}

At any moment in time the observer will receive a transmitted spectrum
that propagated through a medium under non equilibrium conditions.  In
fact, at any given radius, a condition of photoionization equilibrium
cannot be reached, since the time during which the atoms and ions are
subject to the ionizing continuum is much shorter than the
recombination time scale  (the burst duration is assumed to be 100
seconds). In addition, the ionization parameter $\xi\equiv{L}/(nr^2)$
is a function of the distance from the source. For these reasons, it
is not strictly appropriate to model the observed GRB X-ray spectra
with equilibrium models such as WABS\footnote{The WABS photoionization
opacity model (Morrison \& McCammon 1983) assumes that the absorbing
medium is neutral.} or ABSORI\footnote{The ABSORI model (Done et
al. 1992; Zdziarski et al. 1995) considers a warm absorbing medium, at
a temperature $T$ and illuminated by a steady power-law spectrum with
uniform ionization parameter $\xi$.}  within the XSPEC package (Arnaud
1996).

Despite these considerations, it is clearly an interesting issue
whether the transmitted spectra derived in our simulations look like
spectra from an highly ionized medium or not. In other words: if the
bursts explode inside giant molecular clouds, do we expect to see
signatures of absorption from highly ionized atoms in their X-ray
spectra? For example, the Chandra X-ray spectrum of GRB000210 (Piro et
al. 2002) showed evidence of modest ionization (if at all) in its
X-ray spectrum. Should this be considered an evidence against this
burst taking place inside a dense environment?

\begin{figure}
\psfig{file=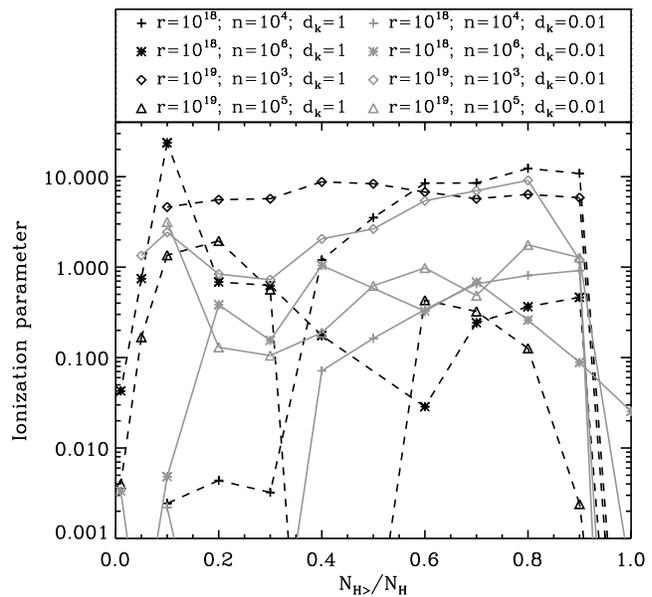,width=0.48\textwidth}
\caption{{Apparent ionization parameter of the absorber as a function of 
the column density decrement for various absorbing clouds. The
apparent ionization parameter is always small, even though the
computed ionization parameter for the cloud outskirts is always very
large, ranging from $\xi=10^7$ to $\xi=10^{10}$.}
\label{fig:xi}}
\end{figure}

\begin{figure}
\psfig{file=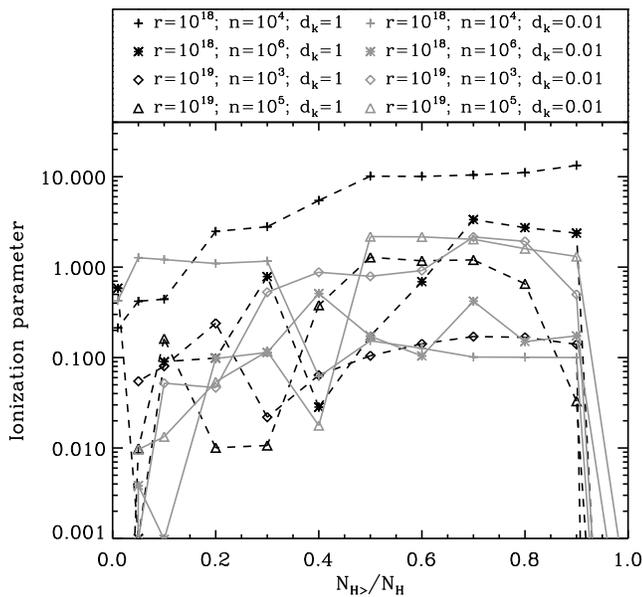,width=0.48\textwidth}
\caption{{Same as Fig.~\ref{fig:xi} but for a spectral index $\alpha=0.5$.}
\label{fig:xia}}
\end{figure}

To understand how much ``ionized'' may look the spectra predicted by
our simulations, we have modelled a set of our synthetic spectra with
the ABSORI model within the XSPEC package (Arnaud 1996). The fits were
performed in the [0.3--20]~keV range, assigning an error bar of $10\%$
to the transmitted spectrum at all frequencies. An infinite resolution
and flat effective area were assumed. The results are shown in
Figs.~\ref{fig:xi} and~\ref{fig:xia} for a set of different geometries
and dust content and two spectral indices. Again, here and in the
following sections, we will consider simulations of an ionizing source
with [1~eV--100~keV] luminosity $L=10^{50}$~erg~s$^{-1}$ turning on at
the centre of a uniform spherical cloud and lasting for $100$~s, after
which is suddenly turned off. It can be seen that, even for big
decrements in the observed absorbing column, the effective ionization
parameter inferred by fitting the spectrum with a warm absorber model
is very small. The presence of dust (dark lines in the figures) has
the effect of making the transmitted spectra seem more ionized, in
particular for the harder spectrum, in which some of the dust grains
are preserved until the end of the simulation (see above). The noise
in the figures is mainly due to the fact that since the error bars
were arbitrarily assumed, there is no point in deriving error bars,
and the fit was converging to different values for different time
shots even if they had very similar $\chi^2$ values.  The real meaning
of the figures is to show that none of the spectra are ever consistent
even with the smallest ionization parameter for the clouds in the
simulation, that was\footnote{For AGN studies, it is more useful
to adopt a different definition $U\equiv{}Q_H/(4\pi\,r^2\,n\,c)$, in
which case we have $U\sim10^3$.}  $\xi=10^7$. The ionization parameter
derived in the fit of a GRB spectrum with a warm absorber should
therefore not be used to infer the physical conditions and geometry of
the cloud that possibly surrounds the source.

\section{Apparent element column densities and relative abundances}

In this last section, before summarizing and discussing our results,
we address the problem of the apparent metallicity of the absorber,
should it be modelled with an absorber model in equilibrium
conditions. Since the ionization time scales for different elements
are different, the residual column density at time $t$ will be
strongly dependent upon the ionizing spectrum and the presence of
dust. For example, Amati et al. (2000) used a cold absorber with
variable metallicity to measure the iron abundance in the X-ray
spectrum of GRB~990705 (see also Lazzati et al. 2001). They find in
this way that iron must be 75 times more abundant than in the sun in
order to reproduce the deep absorption through that was identified
with an iron $K_\alpha$ photoionization edge.

Indeed their measurement is not an absolute measurement, but rather a
comparison between the opacity in the soft X-rays to the opacity at
the frequency of the absorption edge. Since the edge was detected
$\sim 10$ seconds after the GRB onset, the surrounding medium was
heavily ionized. Therefore, the opacity in the soft X-ray band was not
related to the column density of iron nuclei $N_{\rm{Fe}}$, but rather
to the fraction of nuclei with at least one electron
$N_{\rm{Fe}>}$. In addition, the depth of the trough was not compared
to the hydrogen column density $N_H$ but to the absorbing hydrogen
column $N_{H>}$ (or to the oxygen $N_{O>}$, given the BeppoSAX WFC
sensitivity range).

\begin{figure*}
\psfig{file=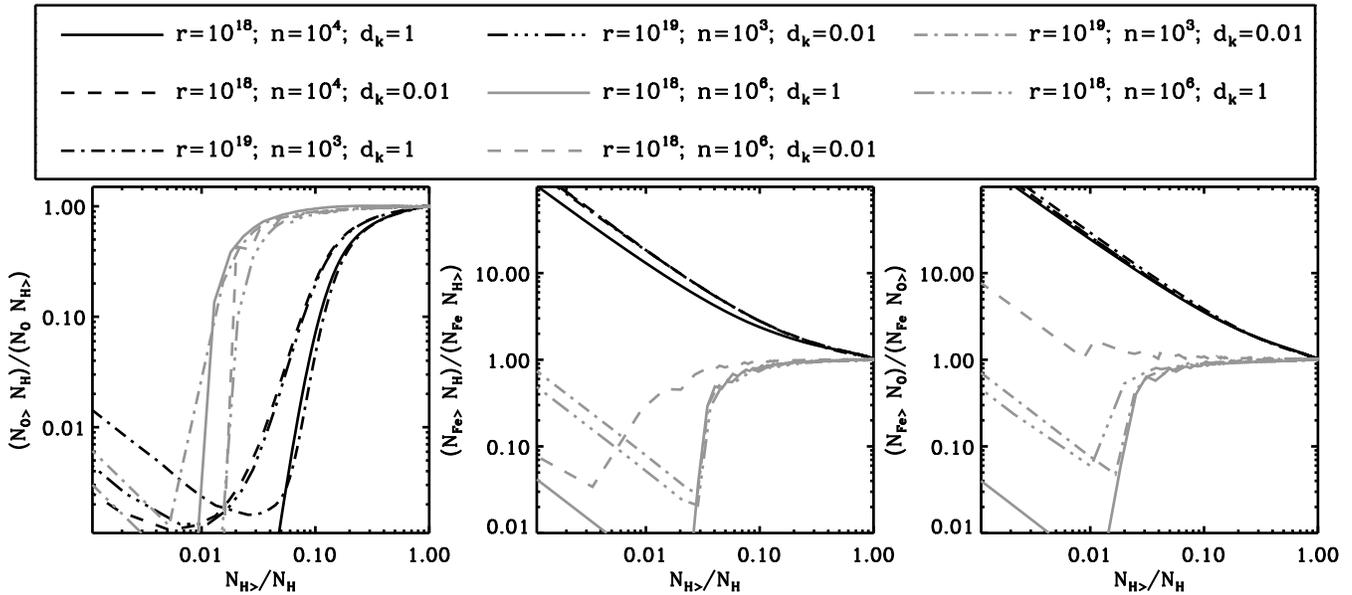,width=\textwidth}
\caption{{Evolution of the apparent abundance of oxygen and iron
versus the evaporation of column. The figure shows, for a set of
geometries and density of the absorbing cloud, the evolution of the
absorbing column of oxygen and iron, the more easy detectable elements
in X-rays. The cloud is centrally illuminated by a source of ionizing
photons with $L=10^{50}$~erg~s$^{-1}$ and $\alpha=0.0$.}
\label{fig:col}}
\end{figure*}

\begin{figure*}
\psfig{file=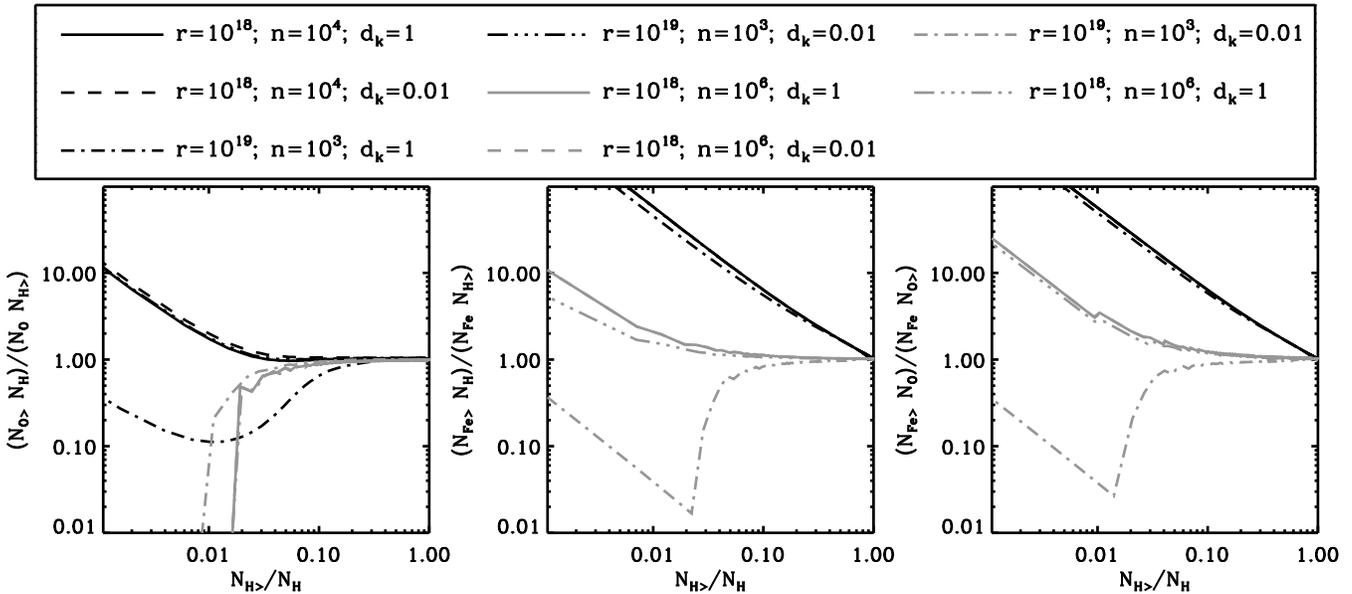,width=\textwidth}
\caption{{Same as Fig.~\ref{fig:col} but for a spectral slope $\alpha=0.5$.}
\label{fig:cola}}
\end{figure*}

The ratio of the column density of ions of a particular element over
the column density of nuclei of the same element in whatever
ionization state (even completely stripped) is highly variable with
time and can be derived from our simulations. Results for hydrogen,
oxygen and iron are shown in Figs.~\ref{fig:col} (for $\alpha=0.0$)
and~\ref{fig:cola} (same but for $\alpha=0.5$). In each figure, the
left panel show the ratio of absorbing oxygen to absorbing hydrogen
column densities (normalized to the initial ``cold'' value) versus the
hydrogen column density at a given time over the initial one. The
central panel shows the same ratios but for the iron-hydrogen
ratio. Finally, the right panel, shows the iron-oxygen ratio. 

The ratio of oxygen to hydrogen with at least one electron decreases
as the global opacity of the medium decreases (i.e. as the time
increases), showing that in most of the cases oxygen is ionized faster
than hydrogen. Only for the ``soft'' spectra of Fig.~\ref{fig:cola}
does the oxygen-to-hydrogen ratio increase with time. A more varied
behavior is displayed by the iron ratio. For clouds with intermediate
initial column densities ($N_H(0)=10^{22}$~cm$^{-2}$; black lines) the
fraction of iron to hydrogen increases with time, while for more
opaque clouds ($N_H(0)=10^{24}$~cm$^{-2}$; gray lines) the
iron-to-hydrogen ratio tends to decrease at the beginning reversing
then the trend at late times (when most of the original opacity has
been burned). This means that an observer will underestimate the
oxygen abundance if fitting the X-ray spectrum with a cold absorber
with variable abundances. On the other hand, the same observer may
either overestimate or underestimate the iron richness depending on
the initial column density of the cloud. In the case of GRB~990705,
Amati et al (2000) compute the iron abundance for a cloud with initial
column few$\times10^{23}$ and find an iron abundance of $\sim75$ when
the column density is still comparable to the initial one. In this
case we can see from Figs~\ref{fig:col} and~\ref{fig:cola} that the
iron richness was overestimated at most by a factor of a few.  Even
though the values of the ratios depend on the presence of dust, an
interesting remark is that the most important parameter seems to be
the initial column density, which determines the general trend of the
evolution.


\section{Discussion and Conclusions}

We have analyzed with a dedicated code the time-dependent absorption
properties of a cloud subject to a strong photon source that sets in
at time $t=0$ at its centre. The novelty of this paper consists in the
fact that the presence of dust and its evaporation is fully taken into
account by the code, allowing us to study if and how the presence of
dust grains can affect the evolution of ionization and vice versa.
This has great importance for variable sources such as GRBs and
Seyfert galaxies, which are sometimes associated with unusual
dust-to-gas ratios (Galama \& Wijers 2001; Stratta et al. 2002), with
variations of the column density (Risaliti et al. 2002; LP02) and with
unusual extinction curves (Maiolino et al. 2001).

In this work we concentrated on the X-ray part of the spectrum, since
the optical and infrared bands were addressed in a companion paper
(Paper II). We study the evolution (evaporation) of the absorption in
X-rays of a uniform cloud with solar metallicity for different dust
contents, initial column density, cloud radius and hardness of the
ionizing spectrum. A source of [1~eV--100~keV] luminosity
$L=10^{50}$~erg~s$^{-1}$ is turned on at time $t=0$ at the centre of
the cloud.

We first concentrated on the effects that the presence of dust can
have on the photoionization rate of metals. We find that, for hard
spectrum sources, the presence of dust grains can have a big effect on
the photoionization of metals (Fig.~\ref{fig:dnd}). In fact, if the
soft photon flux is not large enough to destroy the dust grain by
thermal sublimation, the surviving grains will effectively
shield the entrained metals from photoionization (Fig~\ref{fig:sef}),
allowing them to survive in their neutral state for a long time
(Fig~\ref{fig:sim1}). In the more extreme cases, it is possible that
the ionizing flux completely ionizes all the atoms in the gaseous
phase, leaving behind preferentially large graphite grains, that can
contribute a sizable amount of absorption both at optical and X-ray
wavelengths.  The final conditions of the cloud depend on both the
spectrum, luminosity and variability of the ionizing source as well as
on the cloud size and column density (Fig~\ref{fig:sim1},
\ref{fig:var}, \ref{fig:sima}).

We then study two issues related to observations and their
modelling. It is in fact customary to model the absorption in the soft
X-ray regime by means of equilibrium model for the absorbing
medium. In the case we discuss, equilibrium is far from being reached,
nevertheless it is useful, in order to exploit real observations, to
understand if the predicted spectra look similar to those in
equilibrium and how the column density of the absorbing material that
would be measured does compare to the real one. We discuss the results
of a fit with a cold absorber to our time dependent absorbed spectra
(Fig~\ref{fig:nhs}, \ref{fig:nhsa}), and then show that a condition in
which the absorber seems warm is never reached (Fig~\ref{fig:xi},
\ref{fig:xia}).

We finally consider the observed column densities of the various
elements, which are time dependent, since atoms completely stripped
off their electrons do not contribute to the absorption and are
therefore ``invisible''. We find that these ratios are strong
functions of time and that as a consequence the metallicities inferred
from opacity ratios are not indicative of the pristine composition of
the cloud (Fig~\ref{fig:col}, \ref{fig:cola}).

Present data do not allow for a detailed comparison with our numerical
spectra.  The only evidence of spectral evolution in the early phase
of GRBs and afterglows comes from two detections of decreasing column
density (Connors \& Hueter 1998; Frontera et al. 2000), one variable
absorption edge (Amati et al. 2000) and a puzzling evidence of
variable (but not monotonic) column density (in't Zand et al. 2001).
All these detections belong to the $3\sigma$ realm and do not give
much more information than their mere existence. In the near future,
thanks to the launch of the Swift satellite in 2003, higher quality
spectra will be available, allowing for a more secure detection of
transient features and a more meaningful comparison with
theoretical spectra.

\section*{Acknowledgments}
We are grateful to our referee, Luc Binette, for his careful
report. We thank Luigi Piro for useful discussions and Tomaso Belloni
for technical support. We acknowledge financial support from PPARC
(DL) and from the Harvard Society of Fellows (RP).


\begin{thebibliography}{}

\bibitem{} Amati L., et al., 2000, Science, 290, 953 
\bibitem{} Anders, E., \& Grevesse, N. 1989, Geochim. Cosmochim. Acta,
53, 197
\bibitem{} Arnaud, K. A., 1996, in Jacoby G. H., Barnes J., eds, ASP 
Conf.~Ser. Vol. 101, Astronomical Data Analysis Software and Systems V.
Astron. Soc. Pac., S. Francisco, p. 17
\bibitem{} Connors A., Hueter G.~J., 1998, ApJ,  501, 307
\bibitem{} Done, C., Mulchaey, J.~S., Mushotzky, R.~F., \& Arnaud, 
K.~A.\ 1992, ApJ, 395, 275
\bibitem{} Draine, B. T. \& Hao, L. 2002, ApJ, 569, 780
\bibitem{} Fruchter A., Krolik J. H.  \& Rhoads J. E. 2001, ApJ, 563, 597
\bibitem{} Frontera F., Amati L., Costa E., et al., 2000, ApJS,  127, 59 
\bibitem{} Galama, T. \& Wijers, R. A. M. J. 2001, ApJ, 549, 209
\bibitem{} in't Zand J.~J.~M., Kuiper L., Amati L., et al., 2001, 
	ApJ,  559, 710 
\bibitem{} Laor, A. \& Draine, B. T. 1993, 402, 441
\bibitem{} Lazzati, D., Perna, R., 2002, MNRAS, 330, 383 (LP02)
\bibitem{} Lazzati, D., Perna, R., \& Ghisellini, G. 2001, MNRAS, 325, L19 
\bibitem{} Lazzati, D., Ghisellini, G., Amati, L., Frontera, F., Vietri, 
M., \& Stella, L.\ 2001, ApJ, 556, 471 
\bibitem{} Maiolino R., Marconi A., Salvati M., Risaliti G., Severgnini P., 
	Oliva E., La Franca F., Vanzi L., 2001, A\&A,  365, 28 
\bibitem{} Mathis J.~S., 1990, ARA\&A,  28, 37
\bibitem{} Morrison R., McCammon D., 1983, ApJ, 270, 119
\bibitem{} Perna, R., Loeb, A. 1998, ApJ, 501, 467 
\bibitem{} Perna, R., Raymond, J., \& Loeb, A. 2000, ApJ, 533, 658
\bibitem{} Perna R., Lazzati D., 2002, ApJ, 580, 261 (Paper I)
\bibitem{} Perna R., Lazzati D., Fiore F., ApJ in press 
	(astro-ph/0211235; Paper II)
\bibitem{} Piro, L., Frail, D. A., Gorosabel, J., et al., 2002, ApJ, 57, 680
\bibitem{} Raymond J.~C., 1979, ApJS, 39, 1
\bibitem{} Reilman, R.~F.~\& Manson, S.~T.\ 1979, ApJS, 40, 815 
\bibitem{} Risaliti G., Elvis M., Nicastro F., 2002, ApJ,  571, 234
\bibitem{} Savage B.~D., Mathis J.~S., 1979, ARA\&A,  17, 73
\bibitem{} Schwarz J., 1973, ApJ,  182, 449
\bibitem{} Stratta, G. et al. 2002, ApJ submitted
\bibitem{} Verner, D. A.,  Yakovlev, D. G. 1995, A\&AS, 109, 125
\bibitem{} Waxman, E.; Draine, B. T. 2000, ApJ, 537, 796 
\bibitem{} Zdziarski, A.~A., Johnson, W.~N., Done, C., Smith, D., 
	\& McNaron-Brown, K.\ 1995, ApJ, 438, L63
\end{thebibliography}
\end{document}